\documentstyle[12pt]{article}
\title{Internal environment: What is it like to be a Schr\"odinger cat?}
\author{Hrvoje Nikoli\'c \\
Theoretical Physics Division, Rudjer Bo\v{s}kovi\'{c} Institute, \\
P.O.B. 180, HR-10002 Zagreb, Croatia \\
{\normalsize e-mail: hnikolic@irb.hr} \\
\makebox[1in]{} \\
}
\date{\today}
\begin{document}
\maketitle
\begin{abstract}
%
The possibility of quantum interference of a composite object 
with many internal degrees of freedom is studied, such that the internal degrees play a role 
of an internal environment. 
In particular, if the internal degrees 
have a capacity for an irreversible record of which-path
information, then the internal-environment induced decoherence
prevents external experimentalists from observing interference.
Interference can be observed only if the interfering object is sufficiently isolated from 
the external environment, so that the object cannot record which-path information.
Extrapolation to a hypothetical interference experiment
with a conscious object implies that being a Schr\"odinger cat would be like being an 
ordinary cat living in a box without any information 
about the world external to the box. 
\end{abstract}
\vspace*{0.5cm}
PACS Numbers: 03.65.Ta, 03.65.Yz \newline

\section{Introduction}

Current technology enables quantum interference experiments with large molecules containing
several hundreds of atoms \cite{gerlich} (see also \cite{hornberger,juffmann}). 
A near-future technology might enable  
a quantum interference experiment with viruses \cite{romero-isart}. As the fundamental 
laws of quantum mechanics do not seem to set any boundary on the size or complexity
of objects amenable to interference experiments, the existing rise of technological advance 
rises hope that perhaps one day a sufficiently advanced technology will allow a 
quantum interference experiment similar to the original Schr\"odinger-cat thought 
experiment \cite{schr}, with a conscious object such as a cat or a human.
Even though we are still very far from such a technology, 
the mere idea that such an experiment might be possible in principle rises a provocative
question, of interest to a community much wider than that of quantum physicists.
During an interference experiment with a conscious object, what would be experienced 
by the conscious object herself? Or more operationally, if, after the interference experiment,
we would ask her to report on her experiences during the interference, 
what would she tell us?

Questions on the relation between quantum mechanics and consciousness have also been 
discussed in a classic paper by Wigner \cite{wigner}.
Since wave function determines probability, and since probability changes when 
an observer becomes conscious of a result of observation, Wigner 
argued that there must be an intimate relation between quantum mechanics
and consciousness. Nevertheless, as the origin of consciousness is far from being understood
in physical terms, considerations of consciousness in physical literature are usually
considered controversial. Therefore, 
with a motivation to answer the questions above by using only well-established non-controversial
aspects of quantum mechanics, it is useful to reduce the question to a form
that does not require any reference to consciousness. Hence, 
instead of studying a conscious object that can report on her experiences during the 
interference, we study objects possessing classical information irreversibly recorded
so that it can be read after the interference experiment.
(In this paper, by an ``irreversible'' process we always mean a fundamentally reversible process
which looks irreversible for practical purposes only, due to a very long recurrence time. 
%
If, at this point, it is not clear what do we mean by ``classical'' information in a system ruled by 
the laws of quantum mechanics, 
we note that it will be explained later in the paper.)
For example, the interfering object equipped with an irreversible recorder 
might be a nanorobot many orders of magnitude smaller and simpler than a living cat,
which would be much more amenable to a real quantum interference experiment. 
If we could perform an interference experiment with such an object, 
what information would be read from its irreversible recorder?  

When the question is formulated in this form, the answer is not difficult to guess.
According to the complementarity principle, 
the irreversibly recorded information should not discern any information
that would be complementary to the wave properties seen in the measured interference.
For example, if the measured interference can be explained by a coherent wave 
which travels along two paths at once, then the irreversibly recorded information should not discern any
which-way information that could tell us the actual path that the object took.
Indeed, in this paper we show that this guess is correct and explain in detail why exactly this is so,
at a level suitable to undergraduate students of physics.

For that purpose, we study generic properties of a many-particle quantum state 
describing a composite object made of a large number of particles. 
After describing general kinematics and dynamics of such objects
in Sec.~\ref{SEC2}, we present a detailed study of their interference properties
in a two-path experiment, first in terms
of wave functions in Sec.~\ref{SEC3}, and then in terms of density matrices in Sec.~\ref{SEC4}.
A qualitative discussion of these results is presented in Sec.~\ref{SEC5},
where we make an extrapolation to a hypothetical interference experiment
with a conscious being.

\section{Composite object with many degrees of freedom}    
\label{SEC2}

Consider a composite object made of a large number $N$ of particles, with masses 
$m_1,\ldots,m_N$ and particle positions ${\bf x}_1,\ldots,{\bf x}_N$, the dynamics of which
is described by a Hamiltonian 
\begin{equation}\label{ham}
H({\bf x}_1,\ldots,{\bf x}_N,{\bf p}_1,\ldots,{\bf p}_N)=
\sum_{a=1}^N \frac{{\bf p}_a^2}{2m_a} +V({\bf x}_1,\ldots,{\bf x}_N) .
\end{equation}
We assume that the object is compact, so that the 
space position of the object as a whole can be well described by the position of its center of mass
\begin{equation}
 {\bf X}= \frac{ \sum_{a=1}^N m_a {\bf x}_a }{ \sum_{a=1}^N m_a } .
\end{equation}
The $N$ particle positions can be expressed as positions relative to the center of mass
\begin{equation}
 {\bf q}_a={\bf x}_a-{\bf X} .
\end{equation}
If the center of mass ${\bf X}$ is taken to be one of the new canonical coordinates,
then other new independent canonical coordinates can be taken to be 
$N-1$ relative positions ${\bf q}_a$, say ${\bf q}_2,\ldots,{\bf q}_N$.
For that purpose, it is convenient (but not necessary) to choose 
${\bf x}_1$ to be a ``central'' particle, i.e. a particle the classical position 
of which does not much differ from the center of mass ${\bf X}$.
In this way the new $N$ canonical coordinates can be chosen to be
\begin{equation}\label{eq4}
 {\bf X}, \;\; q\equiv({\bf q}_2,\ldots,{\bf q}_N) ,
\end{equation}
and the Hamiltonian takes a form $H({\bf X},q,{\bf P},p_q)$. 
(Another convenient choice of coordinates for the $N$-particle
problem are the Jacobi coordinates \cite{greiner}.)
Since ${\bf X}$ represents the position of the compact object as a whole, 
the many-component variable $q$ can be thought of as 
representing the internal degrees of freedom of the composite object.

The analysis above can be applied to both classical and quantum mechanics.
Since we are interested in the quantum case, the time evolution
of the composite system is described by the Schr\"odinger equation
\begin{equation}\label{schr}
 H({\bf X},q,\hat{{\bf P}},\hat{p}_q) \Psi({\bf X},q,t)=i\hbar \frac{\partial\Psi({\bf X},q,t)}{\partial t} . 
\end{equation}

When the number of particles $N$ is very large (say, $N\sim 10^{23}$), the experience teaches us
that such systems usually obey classical laws. Does it mean that the Schr\"odinger equation (\ref{schr})
cannot be applied to such systems? Unfortunately, there is no complete consensus among experts
on that question, so any specific claim on that risks to be somewhat controversial. 
(Various attitudes on that issue can be found in \cite{wheeler}.
A more recent attitude based on quantum-classical hybrids is studied in \cite{hybrid},
the details of which, partially due to our incompetence, will not be discussed here.)
Nevertheless, most working physicists seem to agree that classical mechanics is an approximation of quantum mechanics,
and that this approximation can, in one way or another, be heuristically derived for large $N$, even if a fully rigorous
derivation which would satisfy everybody is not known,
and despite the fact that there are even some exceptions to the rule that systems with large $N$ behave
classically. 
Hence, to avoid controversy as much as possible, in this paper we shall comply with this dominating view 
and take for granted that a macroscopic system obeying classical laws is nothing
but a specific many-particle quantum system, the classical properties of which are encoded
in the shape of the wave function. (For example, the fact that a classical object has a well defined
position in space corresponds to a wave function narrow in space.)
In particular, we shall take for granted that a classical object 
which can encode classical information can do that due to a specific form of the wave function, 
and for that purpose we will not need to know what the explicit form of such a wave function is.  

The dynamics described by (\ref{schr}) is too general for making useful predictions.
To further constrain the dynamics, we assume that the total Hamiltonian
has the form
\begin{equation}\label{separ}
H({\bf X},q,{\bf P},p_q) = H_0({\bf X},{\bf P}) + H_q(q,p_q) ,
\end{equation}
where the first term $H_0$ describes dynamics of the object as a whole, while
the second term $H_q$ describes dynamics of the internal degrees of freedom.
The Hamiltonian (\ref{separ}) corresponds to the assumption that  
there is no direct interaction between ${\bf X}$ and $q$.
Many actual systems satisfy this assumption as a very good approximation. 
Eq.~(\ref{separ}) implies that the state $\Psi({\bf X},q,t)$ satisfying 
(\ref{schr}) can be prepared in a product state of the form
\begin{equation}\label{prod_state}
 \Psi({\bf X},q,t)=\psi({\bf X},t)\Phi(q,t).
\end{equation}

Since $\Phi(q,t)$ describes a large number of degrees of freedom, we 
consider a configuration in which the state $\Phi(q,t)$ possesses 
some classical macroscopic properties.
In particular, we assume that it encodes some classical information, 
the evolution of which may be considered irreversible at the macroscopic level.

Furthermore, even though (\ref{separ}) does not contain a direct interaction between
${\bf X}$-degrees and $q$-degrees, it does not necessarily mean that $\psi({\bf X},t)$ and $\Phi(q,t)$ 
in (\ref{prod_state}) are uncorrelated. We assume that the composite object interacts with 
an external environment, on which both $\psi({\bf X},t)$ and $\Phi(q,t)$ may depend.
Hence, the correlation between $\psi({\bf X},t)$ and $\Phi(q,t)$ may be established 
by the interaction with the common external environment. In particular, 
one possibility is that the interaction with the external environment is such that 
the internal $q$-degrees are able to perform a classical 
record of some properties of the external environment, which means that {\em the $q$-degrees
may encode a classical which-path information}.
Such a record has to survive for a long time so that it looks irreversible 
for practical purposes, despite the fact that the true evolution of the whole system 
is a unitary reversible quantum evolution.  
For simplicity, the external environment is treated as a fixed non-dynamical entity, so that
its influence can be described by the potential $V({\bf x}_1,\ldots,{\bf x}_N)$ in (\ref{ham}).

The classical 
irreversibly looking 
behavior of the large number of quantum $q$-degrees makes them
very similar to an external quantum environment widely discussed in the context 
of quantum decoherence \cite{decoh1,decoh2}. The only difference is that the $q$-degrees
are a part of the compact composite object itself. Therefore it is useful to think of
the $q$-degrees as a sort of {\it internal environment} associated with the object.
As a concrete example, the role of internal environments has been studied 
in the context of interference experiments with complex molecules \cite{gerlich,arndt}.
Finally, let us briefly summarize this section by repeating that the whole system we consider
consists of two mayor subsystems: a composite object and its external environment.
The composite object is conceptually further divided into two minor subsystems in (\ref{eq4}),
namely the position ${\bf X}$ of the object as a whole, and the internal environment
describing all the other internal $q$-degrees of the composite object.
All these subsystems are fundamentally quantum, even when, due to a large number of particles, 
some properties of those can approximately be described by classical laws.

\section{Two-path experiment and interference}
\label{SEC3}

It this section we consider an external-environment configuration in which the composite object
may travel along two alternative paths, referred to as path $A$ and path $B$.

Consider first a configuration in which the path $B$ is blocked, so that the 
composite object may travel only along the path $A$. In this case 
the state (\ref{prod_state}) is a product state of the form
\begin{equation}\label{prod_state_A}
 \Psi_A({\bf X},q,t)=\psi_A({\bf X},t)\Phi_A(q,t).
\end{equation}
The corresponding probability density is
\begin{eqnarray}
 \rho_A({\bf X},q,t) & = & \Psi_A^*({\bf X},q,t)\Psi_A({\bf X},q,t)
\nonumber \\
 & = & |\psi_A({\bf X},t)|^2 |\Phi_A(q,t)|^2 .
\end{eqnarray}
If the position ${\bf X}$ of the compact object as a whole is measured
at time $t$ after a travel along the path $A$, then the probability of a given value of ${\bf X}$
is obtained by averaging over all possible values of $q$
\begin{equation}
 \tilde{\rho}_A({\bf X},t)=\int dq \, \rho_A({\bf X},q,t)= |\psi_A({\bf X},t)|^2 ,
\end{equation}
where
\begin{equation}
 dq\equiv d^3q_2\cdots d^3q_N.
\end{equation}

Alternatively, if the path $A$ is blocked so that the 
composite object may travel only along the path $B$, then instead of (\ref{prod_state_A})
we have
\begin{equation}\label{prod_state_B}
 \Psi_B({\bf X},q,t)=\psi_B({\bf X},t)\Phi_B(q,t) ,
\end{equation}
which leads to
\begin{equation}
 \rho_B({\bf X},q,t)= |\psi_B({\bf X},t)|^2 |\Phi_B(q,t)|^2 ,
\end{equation}
\begin{equation}
 \tilde{\rho}_B({\bf X},t)= |\psi_B({\bf X},t)|^2 .
\end{equation}

The most interesting situation, of course, is when neither of the paths is blocked, so that
the object may travel along any of the paths. In this case we have a superposition
\begin{equation}\label{prod_state_AB}
 \Psi({\bf X},q,t)=\frac{\Psi_A({\bf X},q,t)+\Psi_B({\bf X},q,t)}{\sqrt{2}} , 
\end{equation}
so the probability density is
\begin{eqnarray}\label{interf1}
 \rho({\bf X},q,t) & = & \Psi^*({\bf X},q,t)\Psi({\bf X},q,t)
\nonumber \\
& = & \frac{1}{2}[\rho_A({\bf X},q,t)+\rho_B({\bf X},q,t)] 
\nonumber \\
& & + {\rm Re}\,\Psi_A^*({\bf X},q,t)\Psi_B({\bf X},q,t).
\end{eqnarray}
If only the position ${\bf X}$ of the compact object as a whole is measured, then 
averaging over $q$ gives
\begin{eqnarray}\label{interf2}
 \tilde{\rho}({\bf X},t) & = & \int dq \, \rho({\bf X},q,t)
\nonumber \\
& = & \frac{1}{2}[\tilde{\rho}_A({\bf X},t)+\tilde{\rho}_B({\bf X},t)] 
\nonumber \\
& & + \int dq \,{\rm Re}\,\Psi_A^*({\bf X},q,t)\Psi_B({\bf X},q,t) .
\end{eqnarray}
The crucial term in both (\ref{interf1}) and (\ref{interf2}) is the interference term
proportional to ${\rm Re}\,\Psi_A^*({\bf X},q,t)\Psi_B({\bf X},q,t)$. To understand 
the meaning of this term we study two different cases.

The first case is
\begin{equation}\label{case1}
 \Phi_A(q,t)=\Phi_B(q,t)\equiv \Phi(q,t) .
\end{equation}
In this case $\Phi_A(q,t)$ and $\Phi_B(q,t)$ encode the same classical information,
which means that the internal degrees cannot distinguish a travel along path $A$
from a travel along path $B$. In practice this means that the internal degrees
must be isolated from the influence of the external environment, which is actually 
very difficult to achieve in practice when the number of internal degrees of freedom is large.
Nevertheless, the known fundamental laws of physics do not seem to forbid it in principle,
so achieving (\ref{case1}) seems to be a matter of technology. 
So if, with a sufficiently advanced technology,
we can satisfy the condition (\ref{case1}), then (\ref{interf1}) and (\ref{interf2}) reduce to
\begin{equation}\label{rho_int1}
 \rho({\bf X},q,t) = \tilde{\rho}_{\rm interf}({\bf X},t) \, |\Phi(q,t)|^2 ,
\end{equation}
\begin{equation}\label{rho_int2}
 \tilde{\rho}({\bf X},t) = \tilde{\rho}_{\rm interf}({\bf X},t) ,
\end{equation}
respectively, where
\begin{eqnarray}\label{interf_x}
\tilde{\rho}_{\rm interf}({\bf X},t)= \frac{1}{2}[\tilde{\rho}_A({\bf X},t)+\tilde{\rho}_B({\bf X},t)]
\nonumber \\
+ {\rm Re}\,\psi_A^*({\bf X},t)\psi_B({\bf X},t) .  
\end{eqnarray}
The second term in (\ref{interf_x}) is the ${\bf X}$-position interference term,
so both (\ref{rho_int1}) and (\ref{rho_int2}) show that the measurement of  
position ${\bf X}$ of the composite object as a whole leads to a characteristic 
quantum interference pattern. 

The second case is when $\Phi_A(q,t)$ is macroscopically distinct from 
$\Phi_B(q,t)$. This means not only that $\Phi_A(q,t)\neq \Phi_B(q,t)$, but
also that the corresponding probability densities 
\begin{equation}
p_A(q,t)=|\Phi_A(q,t)|^2 , \;\;\;
p_B(q,t)=|\Phi_B(q,t)|^2 ,
\end{equation}
have a negligible overlap, i.e. that
\begin{equation}\label{no-overlap}
 p_A(q,t)\, p_B(q,t) \simeq 0
\end{equation}
for almost all values of $q$ (except perhaps at a set of measure zero).
In terms of wave functions, the requirement (\ref{no-overlap}) can also be written as
\begin{equation}\label{no-overlap2}
 \Phi_A^*(q,t)\, \Phi_B(q,t) \simeq 0 .
\end{equation}
This case is interesting because when $\Phi_A(q,t)$ and $\Phi_B(q,t)$
are macroscopically distinct, then they may encode different classical information.
In particular, they may record classical information about two different paths
along which the composite object may travel. When the number of $q$-degrees of freedom is large,
then such a situation is not difficult to achieve in practice; indeed, 
this is what happens in almost all cases of real experiments.
When the condition (\ref{no-overlap2}) is satisfied, then the interference term in 
(\ref{interf1}) and (\ref{interf2}) is negligible, so (\ref{interf1}) and (\ref{interf2}) 
reduce to
\begin{equation}\label{rho_no-int1}
 \rho({\bf X},q,t) \simeq \frac{1}{2}[\tilde{\rho}_A({\bf X},t)\,p_A(q,t)
+\tilde{\rho}_B({\bf X},t)\,p_B(q,t)],
\end{equation}
\begin{equation}\label{rho_no-int2}
 \tilde{\rho}({\bf X},t) \simeq \frac{1}{2}[\tilde{\rho}_A({\bf X},t)+\tilde{\rho}_B({\bf X},t)] .
\end{equation}
Both (\ref{rho_no-int1}) and (\ref{rho_no-int2}) show the absence of interference.
Hence, when the classical which-path information is encoded in the internal degrees of freedom 
of the composite object, then the measurement of position ${\bf X}$ of the composite object as a whole
cannot reveal quantum interference.

Note that (\ref{rho_no-int2}) would also be obtained by a weaker and more general requirement that 
$\Phi_A(q,t)$ is merely approximately orthogonal to $\Phi_B(q,t)$, without insisting that 
they are macroscopically distinct. Indeed, states which are merely orthogonal 
(without being macroscopically distinct) can also encode different information.
However, they cannot encode different {\em classical} information.
Hence the macroscopically distinct states are needed if the final goal is to understand 
how a macroscopic ``classical'' being (such as a cat) would experience 
an attempt of an interference experiment.   

Finally, it should be noted that many similar experiments in which interference or which-path 
information is observed have actually been performed in laboratories (see e.g. \cite{greenberger,scully}),
with results in agreement with the theory presented above. Such experiments can even perform 
a simultaneous (but imperfect) measurement of both interference {\em and} which-path information,
corresponding to the case in which $\Phi_A$ and $\Phi_B$ have a non-negligible overlap but are not equal,
which interpolates between the two different cases we discussed above.
However, in our case the $q$-degrees of freedom
which record which-path information are internal degrees, while in the experiments actually performed so far
the degrees of freedom that recorded which-path information were external degrees. It is still a
challenge for experimentalists to construct a setup in which one might observe interference 
of a relatively complex object, say a nanorobot, the internal degrees of freedom 
have a capacity for an irreversible record of information.

\section{Density matrix and decoherence}
\label{SEC4}

The results of Sec.~\ref{SEC3} can also be rewritten in the language  
of density matrices, which is particularly useful if we want to interpret the results 
in terms of coherence and decoherence \cite{decoh1,decoh2,auletta}.
For notational simplicity we suppress the dependence on time, so that (\ref{prod_state_AB})
can be written as
\begin{equation}\label{prod_state_AB-m}
 |\Psi\rangle=\frac{|\Psi_A\rangle +|\Psi_B\rangle}{\sqrt{2}} , 
\end{equation}
where
\begin{equation}
 |\Psi_A\rangle=|\psi_A\rangle|\Phi_A\rangle , \;\;\; |\Psi_B\rangle=|\psi_B\rangle|\Phi_B\rangle .
\end{equation}
The corresponding density matrix is
\begin{eqnarray}\label{dens_matr}
 \rho & = & |\Psi\rangle\langle\Psi| 
\nonumber \\
& = & \frac{1}{2} [ |\Psi_A\rangle\langle\Psi_A| +|\Psi_B\rangle\langle\Psi_B|
\nonumber \\
& & +|\Psi_A\rangle\langle\Psi_B| +|\Psi_B\rangle\langle\Psi_A| ].
\end{eqnarray}
This density matrix can be viewed as a state describing two subsystems, 
where one subsystem is microscopic and corresponds to the variable ${\bf X}$, while the other
subsystem is macroscopic and corresponds to the internal-environment variable $q$. We are interested in the
reduced density matrix $\tilde{\rho}$ of the microscopic subsystem, obtained as a partial trace
of $\rho$ over the internal environment degrees  
\begin{equation}\label{part_trace}
\tilde{\rho}={\rm Tr}_q\, \rho = \int dq \, \langle q|\rho|q\rangle .
\end{equation}
For example, we have
\begin{eqnarray}
 \langle q|\Psi_A\rangle\langle\Psi_B|q\rangle & = & |\psi_A\rangle\langle\psi_B| \cdot 
\langle q|\Phi_A\rangle\langle\Phi_B|q\rangle 
\nonumber \\
& = & |\psi_A\rangle\langle\psi_B| \cdot \Phi_B^*(q)\, \Phi_A(q) ,
\end{eqnarray}
which implies
\begin{eqnarray}
\int dq\, \langle q|\Psi_A\rangle\langle\Psi_B|q\rangle & = &
 |\psi_A\rangle\langle\psi_B| \cdot \int dq\, \Phi_B^*(q)\, \Phi_A(q) 
\nonumber \\
& = & |\psi_A\rangle\langle\psi_B| \cdot \langle\Phi_B|\Phi_A\rangle .
\end{eqnarray}
In this way (\ref{part_trace}) with (\ref{dens_matr}) gives
\begin{eqnarray}\label{reduced}
\tilde{\rho} & = & \frac{1}{2}[ |\psi_A\rangle\langle\psi_A| +|\psi_B\rangle\langle\psi_B|
\\
& & +|\psi_A\rangle\langle\psi_B|\cdot \langle\Phi_B|\Phi_A\rangle
+|\psi_B\rangle\langle\psi_A|\cdot \langle\Phi_A|\Phi_B\rangle] .
\nonumber
\end{eqnarray}
This expression can be further simplified by using the representation
\begin{equation}
|\psi_A\rangle =\left(
\begin{array}{c}
1 \\
0
\end{array}
\right) , \;\;\;
|\psi_B\rangle =\left(
\begin{array}{c}
0 \\
1
\end{array}
\right) ,
\end{equation}
which leads to
\begin{equation}\label{matrix}
\tilde{\rho}= \frac{1}{2} \left(
\begin{array}{cc}
1 &  \langle\Phi_B|\Phi_A\rangle \\
\langle\Phi_A|\Phi_B\rangle & 1 
\end{array}
\right) .
\end{equation}
This compact expression is very convenient to distinguish the two cases considered in 
Sec.~\ref{SEC3}. 

In the first case, (\ref{case1}) implies $\langle\Phi_A|\Phi_B\rangle=\langle\Phi_B|\Phi_A\rangle=1$, so
(\ref{matrix}) can be written as
\begin{equation}\label{matrix1}
\tilde{\rho}= \frac{1}{2} \left(
\begin{array}{cc}
1 &  1 \\
1 & 1 
\end{array}
\right) =  |\psi\rangle\langle\psi| ,
\end{equation}
where
\begin{equation}\label{psi_1}
 |\psi\rangle=\frac{|\psi_A\rangle +|\psi_B\rangle}{\sqrt{2}} .
\end{equation}
Hence, when the internal environment cannot distinguish the two paths,
then the density matrix (\ref{matrix1}) 
contains the non-diagonal terms, which corresponds 
to the coherent superposition (\ref{psi_1}) responsible for the fact
that interference can be seen in the microscopic subsystem.

In the second case, (\ref{no-overlap2}) implies $\langle\Phi_A|\Phi_B\rangle\simeq 0$ 
and $\langle\Phi_B|\Phi_A\rangle\simeq 0$, so (\ref{matrix}) becomes
\begin{equation}\label{matrix2}
\tilde{\rho}\simeq \frac{1}{2} \left(
\begin{array}{cc}
1 &  0 \\
0 & 1 
\end{array} 
\right) .
\end{equation}
Hence, when the internal environment can distinguish the two paths,
then the density matrix (\ref{matrix2}) is diagonal. 
The absence of the non-diagonal terms corresponds to the absence of interference in the subsystem 
described by the reduced density matrix.
%

Such a disappearance of the non-diagonal terms in the reduced density matrix
is known as {\em decoherence} \cite{decoh1,decoh2}. 
In the literature, decoherence is usually studied as an effect of external
environment, while in our case decoherence is caused by {\em internal} environment.
In general, both external and internal environment may cause decoherence, and some
kind of decoherence is necessary for a record of which-path information.
In our setup, however, the which-path information is recorded by the internal environment.
This means that the which-path information can be recorded
even without the external decoherence (provided that our composite object is appropriately isolated
from the external environment), but not without the internal decoherence.
The difference between internal and external decoherence is discussed also in
\cite{griffiths,gell-mann,omnes}.

\section{Discussion and conclusion}
\label{SEC5}

We have studied interference of a composite object with many internal degrees of freedom,
so that the internal degrees may carry classical information. 
In particular, the internal degrees
may have a capacity for an irreversible record of which-path
information. However, in agreement with well known principles of quantum mechanics
(see e.g. \cite{schumacher}), 
our analysis shows that it is not possible to have both interference and 
which-path information. Or more generally, interference is not possible
if the state of internal degrees corresponding to the travel along one path is 
macroscopically distinct from that corresponding to the travel along the other path.
To achieve visible interference, it is necessary to isolate the internal degrees
from the external environment, so that the internal degrees cannot 
get classical information about the path of travel.

These results can also be extrapolated to a hypothetical interference experiment 
with a conscious being, such as a cat or a human. Presumably, consciousness
requires a possession of classical information in the brain 
(see also \cite{tegmark}), so our results on classical information
encoded in internal degrees have direct consequences on the state of consciousness.     
Research in cognitive sciences \cite{solomon,oregan,danko} shows that brain and
consciousness cannot work properly without interaction with the environment.
Since we want an experiment with a conscious being whose consciousness 
works properly, we can put the being into a closed box within which all
supplies needed for normal conscious life are present. The box is supposed to 
serve as a shield from the influence of {\em external} environment, i.e.
environment outside of the box. In this way the whole box, together 
with the conscious being inside it, is considered to be a composite object
that suffers interference. In the interference experiment itself, the external observer 
measures only the position of the box as a whole. 
With such a hypothetical experiment, we can definitely answer the question
posed in the title of this paper, namely: ``What is it like to be a Schr\"odinger cat?''

Before answering the question, to avoid possible confusion let us first clarify the terminology.
In the original thought experiment \cite{schr}, the Schr\"odinger cat
had a very narrow meaning, corresponding to a cat which was supposed to be in a 
superposition of a dead and alive cat. In the modern terminology the notion of a ``Schr\"odinger cat''
has a much wider meaning, referring to any superposition of macroscopically distinct states of matter or light.
In the present paper (the title of which paraphrases the title of a famous paper on 
philosophy of mind \cite{nagel}) the meaning of the notion of ``Schr\"odinger cat'' is somewhere in between.  
By a ``Schr\"odinger cat'' we mean a superposition of two macroscopically distinct states of a conscious being,
which is more narrow than its modern meaning,
but wider than meaning in the original thought experiment.
Now the answer to the question is very simple.
%
In order for interference to be seen by the 
external observer, the box must be almost perfectly isolated from the influence
of the external environment, so that the conscious being in the box cannot have any 
information about the external world. Therefore, even though the wave function of the conscious being
would travel along both paths (as would be demonstrated by the measurement of interference),
the conscious being would experience nothing unusual because she would not even know
that she travels along some of the paths. Hence, {\em being a Schr\"odinger cat would be like being an 
ordinary cat living in a box without any information 
about the world external to the box}. 

Alternatively, if the box does not 
provide such an almost perfect isolation from the influence of external environment, 
then the conscious being may know the path along which she travels, but then interference
cannot be observed. This, indeed, is what happens in everyday macroscopic phenomena
and experimentally cannot be distinguished from a classical situation in which the 
being travels only along one of the paths.       

Finally, let us note that these results also help to resolve the Wigner's friend problem \cite{wigner}.
Instead of considering a cat, Wigner considers his friend who first performs an observation
and then reports the result to Wigner. When does the wave function suffer a change,
only when Wigner's friend reports the result to Wigner, or even before, 
e.g. when Wigner's friend makes the observation?
To answer that question posed in \cite{wigner}, we think of Wigner as an external observer
and Wigner's friend as a part of a composite quantum system with many internal degrees of freedom. 
Since the number of internal degrees is large, most of them cannot be observed in practice.
These degrees may carry information needed to observe quantum interference,
implying that interference cannot be observed in practice when these degrees are unobservable in practice.
In this way, interference may effectively disappear for practical purposes, despite the fact that
all degrees of a closed system do carry all interference effects at the fundamental level.  
Hence the relevant change of the wave function can be identified with the 
effective
disappearance of interference, and the ability to observe can be identified with the ability
to 
effectively
destroy the interference. 
In this way, even without an explicit reference to consciousness,
it should be clear that Wigner's friend 
effectively
destroys the interference, i.e. that the wave function
suffers a relevant change even before Wigner's friend reports the result to Wigner.    

\section*{Acknowledgments}

%
%
The author is grateful to Danko Nikoli\'c for discussions 
and useful comments on the manuscript,
and to anonymous referees for various suggestions to improve the paper.
This work was supported by the Ministry of Science of the
Republic of Croatia. 

\end{document}